\documentclass[aps,preprint]{revtex4}
\usepackage[dvips]{graphicx}
\usepackage{amsmath, amssymb}

\begin{document}
\title{Orbital magnetic susceptibility of zigzag carbon nanobelts: a tight-binding study}

\author{Norio Inui}

\address{Graduate School of Engineering, 
University of Hyogo, 
Himeji, Hyogo 671-2280, Japan}
\begin{abstract}
The magnetic properties of a circular graphene nanoribbon (carbon belt) in a magnetic field parallel to its central axis is studied using a tight-binding model. 
Orbital magnetic susceptibility is calculated using an analytical expression of the energy eigenvalues as a function of the magnetic flux density for any size, and its temperature dependence is considered. In the absence of electron hopping parallel to the magnetic field, the orbital magnetic susceptibility diverges at absolute zero if the chemical potential is zero and the number of atoms is a multiple of four. 
As the temperature increases, the magnitude of susceptibility decreases  according to the power law, whose exponent depends on the size. In the presence of electron hopping parallel to the magnetic field, the divergence of the susceptibility near absolute zero disappears, and the sign changes with the transfer integral parallel to  the magnetic field and the temperature.
\end{abstract}

\maketitle

\section{Introduction}\label{S1}
Low-dimensional materials with honeycomb structures such as C$_{60}$ 
\cite{Elser1987,Coffey1992},
carbon nanotube \cite{Ajiki1993,Lu1995,Chen2005}, and graphene \cite{Sepioni2010,Ortega2013}, have attracted attention. Recently, 
a circular graphene nanoribbon, called a carbon nanobelt 
\cite{Povie2017}, has been added to allotropes of carbon. 
As defined by Itami et al. \cite{Segawa2016,Itami2023}, a carbon nanobelt is a ring-shaped segment of a carbon nanotube, and the cleavage of at least two C-C bonds is necessary  to open it.
The first carbon nanobelt is a segment of (6,6) carbon nanotubes and consists of 12 benzene rings. Longer and complex carbon nanobelts can be achieved using current synthesis techniques \cite{Li2021,Wang2023,Wang2024}.
Although many physical properties of carbon nanobelts remain unknown, researchers have reported high electric conductance between the tip of a scanning tunneling microscope and carbon nanobelts; this excellent property is expected to benefit electric applications
\cite{Li2022}.

A carbon nanotube can be regarded as a graphene sheet rolled into a tube \cite{Li2013}.
Similarly, a carbon nanobelt can be regarded as a graphene nanoribbon that can be seamlessly bent into cylindrical shapes.
Graphene exhibits unique magnetic properties not observed in other materials \cite{Goerbig2011}.
For example, divergent orbital diamagnetism at the Dirac point has been theoretically predicted and was 
recently detected \cite{Bustamante2021}. The magnetic properties of graphene ribbons are primarily determined by two factors: 
orbital magnetization and spin angular momentum \cite{Sepioni2010}. This diamagnetic property originates from a previous contribution. Unlike the spin, the orbital magnetization of a graphene sheet is generated by a global electric current, which depends on the shape of the sheet. In this study, we focus on the magnetization of carbon nanobelts caused by the orbital current circulating around them.

 Extensive theoretical studies on the orbital magnetization of graphene disks have revealed that it strongly depends on their size and type of edges, temperature, and defects \cite{Ezawa2007,Liu2008,Ominato2013,Deyo2021}. However, little is known about the magnetic properties of carbon nanobelts. Thus, we introduce a tight-binding model for carbon nanobelts and investigate the dependence of the magnetic susceptibility on their size and temperature. The advantage of using a tight-binding model over other theoretical methods, such as density functional theory \cite{Shi2020,Wu2013,Negrin2023}, is that the energy eigenvalue can often be expressed in analytical forms for arbitrary sizes. The magnetic susceptibility is obtained by calculating the second derivative of the free energy \cite{Ominato2013}. Thus, the dependence of the magnetic susceptibility on the size can be precisely  discussed.

The remainder of this paper is organized as follows:
In Section 2, we introduce a tight-binding model for a carbon nanobelt based on cyclacene, which is the first proposed carbon nanobelt \cite{Turker2004}. 
Although cyclacene has not yet been synthesized, its structure is the simplest among carbon nanobelts \cite{Gleiter2009}. Cyclacene can be considered as two interacting one-dimensional atomic arrays,  hereafter referred to as rings. Therefore, we address the energy eigenvalue problem of a ring  and explain its size dependence. 
In Section 3, 
the dependence of the magnetic susceptibility of the ring on its size and temperature is discussed.
In Section 4, the magnetic properties of the carbon nanobelts, 
 hereafter referred to as belts,  are investigated.
In particular, the difference in  the magnetic susceptibility between the rings and the belts is discussed. In Section 5, the relations between the change in the energy level by applying a magnetic field and the magnetic properties  is summarized. 

\section{Tight-binding model of a ring}
The structure of the belt considered in this study consists of  upper and lower rings, as shown in Fig. 1(a),  where the solid circles indicate the atoms.
A uniform magnetic field is applied parallel to the central axis ($z$-axis) of the belts. Figure  1(b) shows the position of atoms in the upper ring projected onto the $xy$-plane. We assume that atoms in a ring exist on a circle. 
The electrons can hop along the circumference 
and  is characterized by the transfer integral $\gamma$. Additionally, 
electrons can hop between rings, whose transfer integral $\gamma^{\prime}$ 
may be different from $\gamma$. Because 
the magnetic response to the these parameter dependencies 
cannot be easily considered simultaneously, we prohibited 
hopping between rings in this section and considered 
the magnetic response of a single ring using a tight-binding model. 

\begin{figure}[h]
\hspace{20mm}
\begin{center}
\includegraphics[bb= 0 0 290 170 , origin = c, clip,scale=1.0]{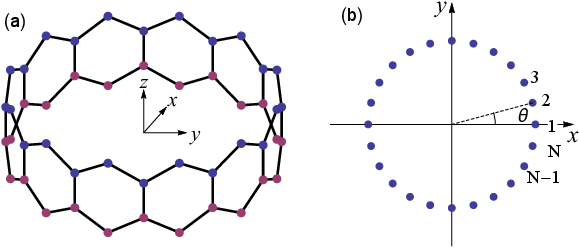}
\end{center}
\caption{(a) Configuration of a belt fused 12 benzene rings. Magnetic field is applied parallel to $z$-axis. (b) Labeling and positions of upper sites indicated by blue circles, where $\theta$ = $2\pi/N$.}
\end{figure}

Let the number of sites (atoms) in  the ring be $N$. 
The radius of a ring, whose nearest distance between sites in the $xy$-plane is $a$ is expressed as
\begin{eqnarray}
R_{N} &=& \frac{a}{2\sin(\frac{\pi}{N})}.
\end{eqnarray}
Each site in the ring is labeled with an integer $j$ $\in$ $\{1,2,\ldots,N\}$ as shown in Fig. 1(b), and the position of the $j$-th site on the $xy$-plane $\vec{r}_{j}$ = $(x_{j},y_{j})$ is defined by
\begin{eqnarray}
\left \{
\begin{tabular}{l}
$x_{j}$ = $R_{N}\cos\left(\frac{2\pi (j-1)}{N}\right)$, \\
$y_{j}$ = $R_{N}\sin\left(\frac{2\pi (j-1)}{N}\right)$.
\end{tabular}
\right.
\label{xy}
\end{eqnarray}

The Hamiltonian of the tight-binding model  \cite{Ezawa2007}, in which the contribution of spin is neglected, is defined by
\begin{eqnarray}
H_{\mbox{\tiny Ring}} &=& -\gamma\sum_{j=1}^{N}
(e^{i\phi_{j}}c_{j+1}^{\dag}c_{j}+
e^{-i\phi_{j}}c_{j}^{\dag}c_{j+1}).
\label{Hamiltonian}
\end{eqnarray}
where  the operator $c_{j}^{\dag}$ creates an electron at the $j$-th site, and $c_{j}$ annihilates an electron at the $j$-th site. 
The periodic condition requires $c_{N+1}^{\dag}$ = $c_{1}^{\dag}$ and $c_{N+1}$ = $c_{1}$.
The effect of the magnetic field in the Hamiltonian is expressed through  the Peierls phase $\phi_{j}$ between the $j$-th site and the $(j+1)$-th site defined by
\begin{eqnarray}
\phi_{j} &=& -\frac{e}{\hbar}\int_{\vec{r}_{j}}^{\vec{r}_{j+1}}
d\vec{r}\cdot\vec{A},
\label{Peierls}
\end{eqnarray}
where $\vec{A}$ is the vector potential and is related to the applied magnetic flux density $\vec{B}$ in the form of $\vec{B}$ = $\nabla \times \vec{A}$. Using the Landau gauge, the vector potential can be expressed as
\begin{eqnarray}
\vec{A}&=& (0,Bx,0).
\label{vecA}
\end{eqnarray}
In the continuous limit of $a$ $\rightarrow$ 0, the Hamiltonian (\ref{Hamiltonian}) can be  approximately regarded as $(\hat{p}+e\vec{A})^2/2m$, where $\hat{p}$ and $m$ are the momentum operator and effective mass related to $\gamma$, respectively \cite{Boykin2001,Inui2023}. Combining Eq. (\ref{vecA}) with Eq. (\ref{Peierls})  leads to 
\begin{eqnarray}
\phi_{j} &=& \frac{eB}{2\hbar}(x_{j+1}+x_{j})(y_{j}-y_{j+1}), \\
&=& -\alpha 
\sin\left(\frac{2\pi}{N}\right)
\cos^2\left(\frac{(2j-1)\pi}{N}\right),
\end{eqnarray}
where $\alpha$ is a dimensionless parameter defined by
\begin{eqnarray}
\alpha &=& \frac{eR_{N}^2B}{\hbar}.
\end{eqnarray}
If the magnetic flux density through the ring is defined by $\phi$ $\equiv$ 
$\pi R_{N}^2 B$, then the parameter $\alpha$ is expressed as 
$\phi/\phi_{0}$ where $\phi_{0}$ $\equiv$ $h$/(2$e$) is magnetic flux quantum.

The energy eigenvalue $\epsilon_{n,N}(\alpha) $ of the Hamiltonian $H_{\mbox{\tiny Ring}}$ with quantum number $n$ = 1, 2, $\ldots$ $N$ is given by
\begin{eqnarray}
\epsilon_{n,N}(\alpha) &=& -2\gamma\cos
\left[
\frac{2n\pi}{N}
+\frac{\alpha}{2}
\sin\left( \frac{2\pi}{N}\right) 
\right].
\label{eigenvalue1}
\end{eqnarray}

The $j$-th component of the eigenstate $\psi_{n}$ corresponding to $\epsilon_{n,N}(\alpha)$ is given by
\begin{eqnarray}
\psi_{n,j} &=& \frac{1}{\sqrt{N}}e^{i\theta_{n,j}}, 
\end{eqnarray}
where
\begin{eqnarray}
\theta_{n,j} &=&
\frac{2nj\pi}{N}
-\frac{\alpha}{4}
\left[
\sin
\left( 
\frac{4\pi}{N}
\right)
+
\sin
\left(
\frac{4(j-1)\pi}{N}
\right) 
\right].
\end{eqnarray}
The combination of the eigenvalue and eigenstate shown above satisfies the Schr\"{o}dinger equation $H_{\mbox{\tiny Ring}}\psi_{n}$ 
= $\epsilon_{n,N}\psi_{n}$.
Applying the Hamiltonian on the eigenstate, we have
\begin{eqnarray}
[H_{\mbox{\tiny Ring}}\psi_{n}]_{j}&=&-\gamma(e^{i\phi_{j-1}} \psi_{n,j-1}
+e^{-i\phi_{j}}\psi_{n,j+1}),
\label{Hpsi1}
\end{eqnarray}
where $[.]_{j}$ denotes the $j$-th component of $H_{\mbox{\tiny Ring}}\psi_{n}$.
The eigenstates at the sites $j+1$ and $j-1$ can be expressed as
\begin{eqnarray}
\psi_{n,j+1} &=&e^{i\delta_{n,j}} \psi_{n,j}, \label{shift1}\\
\psi_{n,j-1} &=&e^{-i\delta_{n,j-1}} \psi_{n,j},\label{shift2}
\end{eqnarray}
where
\begin{eqnarray}
\delta_{n,j} = \frac{2n\pi}{N}-\frac{\alpha}{2}
\sin\left(\frac{2\pi}{N}\right)\cos\left(\frac{2(2j-1)\pi}{N} \right).
\end{eqnarray}
Combining Eqs. (\ref{Hpsi1}), (\ref{shift1}), and (\ref{shift2}), we obtain
\begin{eqnarray}
[H_{\mbox{\tiny Ring}}\psi_{n}]_{j}&=&-\gamma\left(e^{-i
\left(
\frac{2n\pi}{N}
+\frac{\alpha}{2}\sin
\left(
\frac{2\pi}{N}
\right) \right)} \psi_{n,j}
+e^{i
\left(
\frac{2n\pi}{N}
+\frac{\alpha}{2}\sin
\left(
\frac{2\pi}{N}
\right) \right)} \psi_{n,j} 
\right),
\\
&=&-2\gamma\cos
\left[
\frac{2n\pi}{N}
+\frac{\alpha}{2}
\sin\left( \frac{2\pi}{N}\right) 
\right]\psi_{n,j}, \\
&=&\epsilon_{n,N}(\alpha)[\psi_{n}]_{j}.
\end{eqnarray}

We define the energy eigenvalue  normalized  by $\gamma$ as $\lambda_{n,N}(\alpha)$ 
($\equiv$ $\epsilon_{n,N}(\alpha)/\gamma$).
Figure 2 shows $\lambda_{n,N}(\alpha)$ for $N$ = 4, 5, and 6 as a function of $\alpha$.
In the absence of the magnetic field, i.e., $\alpha$ = 0, each energy level has a degeneracy of two except for
$n$ = 1 (grand state) and $n$ = $N$. Note that $\lambda_{2,4}(\alpha)$ and $\lambda_{3,4}(\alpha)$ are zero at $\alpha$ = 0. This holds for any $N$ if $N$ mod 4 = 0, 
i.e., if $N$ is a multiple of 4.

\begin{figure}[h]
\hspace{20mm}
\begin{center}
\includegraphics[bb= 0 0 240 500 , origin = c, clip,scale=1.0]{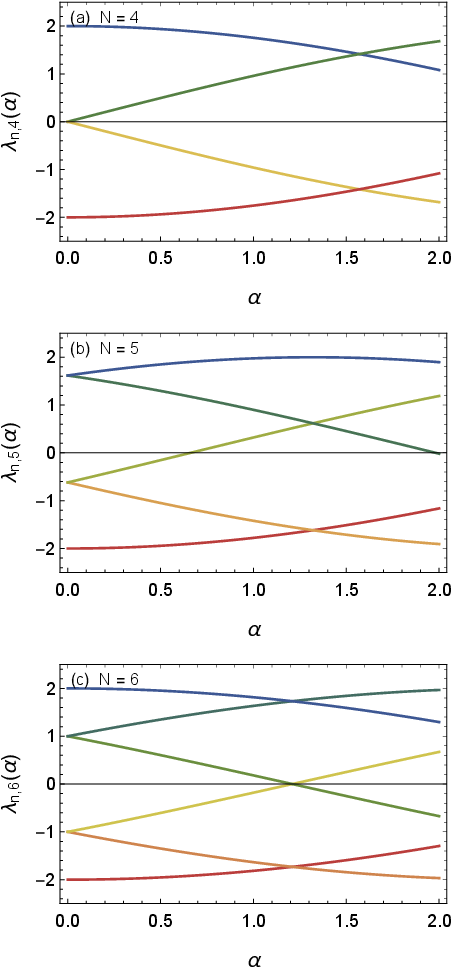}
\end{center}
\caption{Energy eigenvalues of rings normalized by the transfer integral $\gamma$ for $N$ = 4, 5, and 6.}
\end{figure}

The number of six-membered rings in carbon nanobelts 
reported in Ref. \cite{Povie2017} is 12, which is greater than that in the site considered above. 
Although we showed the eigenvalues of the rings with a small $N$ for  visibility, 
the exact eigenvalues can be  obtained for an arbitrary $N$ 
and magnetic flux $B$ using Eq. (\ref{eigenvalue1}).


\section{Magnetic susceptibility of a ring}
\subsection{Relation between magnetic susceptibility and energy eigenvalues}
Magnetic susceptibility is an important parameter that determines the magnetic properties of a ring and depends on the temperature $T$. The temperature dependence is calculated from free energy as follows \cite{Ominato2013}:
\begin{eqnarray}
F_{N}(\alpha,T) &=& -2k_{\mbox{\tiny B}}T
\sum_{n}
\ln[1+e^\frac{\mu-\epsilon_{n,N}(\alpha)}{k_{\mbox{\tiny B}}T}],
\label{FreeEnergy}
\end{eqnarray}
where the factor ``2'' originates from the degree of spin polarization, and $k_{\mbox{\tiny B}}$ and $\mu$ are the Boltzmann constant and  chemical potential, respectively \cite{Deyo2021,Maebuchi2023}.
In the following, we consider only the case of $\mu$ = 0. 
The magnetic susceptibility per unit length \cite{Wakabayashi1999} is given by
\begin{eqnarray}
\chi &=& \left.-\frac{1}{2\pi R_{N}}
\left(\frac{\partial^2 F}{\partial B^{2}} \right)_{T} \right|_{B=0}.
\label{defchi}
\end{eqnarray}
Figure 3 shows the magnetic susceptibility normalized by $\chi_{0}$ $\equiv$ $\gamma e^2 a^3/8\pi \hbar^2$ as a function of 
the normalized temperature defined by $k_{\mbox{\tiny B}}T/\gamma$. The magnetic susceptibility for $N$ = 4 diverges, as $T$ approaches zero.

\begin{figure}[h]
\hspace{20mm}
\begin{center}
\includegraphics[bb= 0 0 260 180 , origin = c, clip,scale=1.0]{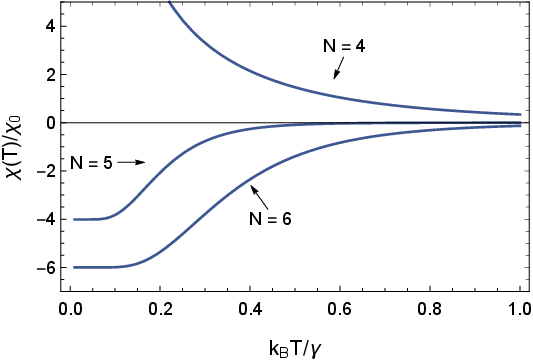}
\end{center}

\caption{Temperature dependence of the  magnetic susceptibilities of rings normalized with $\chi_{0}$ for 
$N$ = 4, 5, and 6. }
\end{figure}

Let the inverse temperature normalized with $k_{\mbox{\tiny B}}/\gamma$ be $\beta$ $\equiv$ $\gamma/k_{\mbox{\tiny B}}T$. Substituting Eq. (\ref{FreeEnergy}) into Eq. (\ref{defchi}), the susceptibility is expressed as
\begin{eqnarray}
\chi &=& -\frac{\chi_{0}}{4\sin^3\left(\frac{\pi}{N}\right)}
\sum_{n=1}^{N}
\left[
4\lambda_{n,N}^{\prime\prime}(0)
G_{1}\left(-\beta \lambda_{n,N}(0)\right) \right. \nonumber \\
&&
\left.
-\beta(\lambda_{n,N}^{\prime}(0))^2
G_{2}\left(-\frac{\beta}{2} \lambda_{n,N}(0)\right)
\right] ,
\label{chiesum1}
\end{eqnarray}
where
\begin{eqnarray}
G_{1}(x) &=& \frac{1}{1+e^{-x}} ,\\
G_{2}(x) &=& \mbox{sech}^2(x) \equiv \frac{4}{(e^{x}+e^{-x})^2}.
\end{eqnarray}
Equation  (\ref{chiesum1}) indicates that the susceptibility is determined from the energy eigenvalues and their derivative up to the second order, and this is  necessary for considering the relation between the susceptibility and the energy eigenvalues.
Let $\xi_{n}$ be $2 \pi n/ N$. Using the expression for eigenvalues Eq. (\ref{eigenvalue1}), we have 
\begin{eqnarray}
\chi &=& 
-\chi_{0}\frac
{\cos^2\left( \frac{\pi}{N}\right)}
{\sin\left( \frac{\pi}{N}\right)} 
\sum_{n=1}^{N}
(\zeta_{n}^{(1)}(\beta)+\zeta_{n}^{(2)}(\beta)),
\end{eqnarray}
where
\begin{eqnarray}
\zeta_{n}^{(1)}(\beta) &=& 
2\cos\left(\xi_{n}\right)
G_{1}\left[2\beta \cos\left(\xi_{n}\right)\right],
\\ 
\zeta_{n}^{(2)}(\beta) &=& 
-\beta\sin^2\left(\xi_{n}\right)
G_{2}\left[\beta\cos\left(\xi_{n}\right)\right].
\end{eqnarray}

\subsection{Magnetic susceptibility at absolute zero}
To consider the magnetic susceptibility at low temperatures,
 two contributions $\zeta_{n}^{(1)}$ 
and $\zeta_{n}^{(2)}$ to susceptibility should be examined separately. The former is related to the second derivative at $\alpha$ = 0, and the latter is related to the first derivative at $\alpha$ = 0. In the limit of $T$ $\rightarrow$ 0 (i.e. $\beta$ $\rightarrow$ $\infty$),
$\zeta_{n}^{(1)}(\alpha,T)$ converges as follows:
\begin{eqnarray}
\lim_{\beta \rightarrow \infty}
\zeta_{n}^{(1)} 
&=& 
\left \{
\begin{tabular}{ll}
$2\cos\left(\xi_{n}\right)$, & $\cos(\xi_{n})$ $\geq$ 0,\\
0, & $\cos(\xi_{n})$ $<$ 0.
\end{tabular}
\right.
\end{eqnarray}
Accordingly, the sign of $\cos(\xi_{n})$ is important.
The conditions of $\cos(\xi_{n})$ = 0 are that $N$ mod 4 = $0$ and $n$ = $N/4$, $3N/4$. This is the same as the condition that the energy eigenvalue is zero in the absence of the magnetic field.
The range where $\cos(\xi_{n})$ is positive is given by
\begin{eqnarray}
J_{+}(N) &=&
\left\{
1 \leq n \leq \frac{N-(N\,\mbox{mod}\,4)}{4}
\right\}
\bigcap
\left\{
\frac{3N+(N\,\mbox{mod}\,4)}{4} \leq n \leq N
\right\}. \nonumber \\
\end{eqnarray} 
The summation of $\cos(\xi_{n})$ over $n$ satisfying $\cos(\xi_{n})$ $>$ 0 is written as
\begin{eqnarray}
\sum_{n \in J_{+}(N)}
\cos\left(\xi_{n}\right)
&=&
\left \{
\begin{tabular}{ll}
$\frac{1}{\tan\left(\frac{\pi}{N}\right)}$,
& $N$ mod 4 = 0, \\
$\frac{1}{2\sin\left(\frac{\pi}{2N}\right)}$,
& $N$ mod 4 = 1, 3, \\
$\frac{1}{\sin\left(\frac{\pi}{N}\right)}$,
& $N$ mod 4 = 2. \\
\end{tabular}
\right.
\end{eqnarray}

We now  consider the second contribution. If $\cos(\xi_{n})$ is not zero,
$\beta\cos(\xi_{n})$ diverges in the limit of $\beta \rightarrow \infty$.
Thus, $G_{2}(\beta\cos(\xi_{n}))$ converges to zero at  absolute zero.
Namely, the second contribution is neglectable near $T$ = 0 if $\cos(\xi_{n})$ is not zero. If $\cos(\xi_{n})$ is zero, $\zeta_{n}^{(2)}$ is expressed as $-\beta$, as $\beta$ approaches $\infty$. 
Therefore, 
if $N$ is a multiple of 4, then 
the contribution of $\zeta_{n}^{(2)}$ dominates 
the temperature dependence near  absolute zero. 
Because the values of $\cos(\xi_{n})$ is zero at  $n$ = $N/4$ and  $3N/4$,
 the susceptibility of carbon nanobelts near $T$ = 0 is expressed as
\begin{eqnarray}
\chi &\approx& 
-\chi_{0}\frac
{\cos^2\left( \frac{\pi}{N}\right)}
{\sin\left( \frac{\pi}{N}\right)} 
\left(
\zeta_{\frac{N}{4}}^{(2)}(\beta)+\zeta_{\frac{3N}{4}}^{(2)}(\beta)
\right), \hspace{3mm}N
 \, \mbox{mod}\, 4 = 0, \\
&=&
2\chi_{0}\frac
{\cos^2\left( \frac{\pi}{N}\right)}
{\sin\left( \frac{\pi}{N}\right)} 
\frac{\gamma}{k_{\mbox{\tiny B}}T},
\hspace{3mm}N \, \mbox{mod}\, 4 = 0.
\end{eqnarray} 
To summarize, if $N$ mod 4 = 0, the susceptibility diverges in proportion to $T^{-1}$ in the limit of $T$ $\rightarrow$ $0$. If $N$ mod 4 $\neq$ 0, the susceptibility at $T$ = 0 is given by 
\begin{eqnarray}
\chi_{T=0}
&=&
\left \{
\begin{tabular}{ll}
$-\chi_{0}\frac{\cos^{2}\left(\frac{\pi}{N}\right)}{\sin\left(\frac{\pi}{N}\right)\sin\left(\frac{\pi}{2N}\right)}$,
& $N$ mod 4 = 1, 3, \\
$-\chi_{0}\frac{2\cos^{2}\left(\frac{\pi}{N}\right)}{\sin^2\left(\frac{\pi}{N}\right)}$,
& $N$ mod 4 = 2. \\
\end{tabular}
\right.
\end{eqnarray}
This asymptotic behavior is valid only when the chemical potential is zero.  More generally, 
$\mu-\epsilon_{n,N}(0)$ determines the difference between divergence and convergence of $\chi$ near $T$ = 0.
For example, the condition that divergence occurs in the limit of $T$ $\rightarrow$ $0$ is replaced with  $\cos(\xi_{n})$ = $\mu$/$\gamma$.


\subsection{Asymptotic behavior of magnetic susceptibility at high temperatures}
The magnetic susceptibility converges to zero as the temperature increases, as shown in Fig. 3. 
We consider the asymptotic behavior of the magnetic susceptibility at high temperatures.
In the limit of $T$ $\rightarrow$ $\infty$, the parameter $\beta$ converges to zero. Thus,
we analyze the Taylor series of $\zeta_{n}$ $\equiv$ $\zeta_{n}^{(1)}$ + $\zeta_{n}^{(2)}$ near $\beta$ = 0. The expressions for the Taylor series of $G_{1}(x)$ and $G_{2}(x)$ are as
follows:
\begin{eqnarray}
G_{1}(x) &=& \frac{1}{2}+\sum_{k=0}^{\infty}a_{2k+1}x^{2k+1},
\label{G1series} \\
G_{2}(x) &=& \sum_{k=0}^{\infty}b_{2k}x^{2k},
\label{G2series}
\end{eqnarray}
where 
\begin{eqnarray}
a_{k} &=&
-\frac{1}{k!}\sum_{j=1}^{k}\sum_{i=1}^{j}
\frac{(-1)^i}{2^{j + 1}} 
\left(
\begin{tabular*}{2.5mm}{c}
\hspace{-2mm}$j$\hspace{-1mm} \\
\hspace{-2mm}$i$\hspace{-1mm}
\end{tabular*}
\right)i^k, \\
b_{2k} &=& (1+2k)4^{k+1}a_{2k+1}.
\label{b2k}
\end{eqnarray}
The details of the calculations for this series expansion is described in Appendix.
Substituting $\beta\cos(\xi_{n})$ into $x$ in the Taylor series in Eqs. (\ref{G1series}) and (\ref{G2series}) yields
\begin{eqnarray}
\zeta_{n} &=& 
\cos\left(\xi_{n}\right)+
\sum_{k=0}^{\infty}
c_{n,k}\beta^{2k+1}, 
\end{eqnarray}
where
\begin{eqnarray}
c_{n,k} &=& 
2a_{2k+1}2^{2k+1}
\cos^{2k+2}\left(\xi_{n}\right)
+
b_{2k}\cos^{2k+2}\left(\xi_{n}\right) 
-
b_{2k}
\cos^{2k}\left(\xi_{n}\right). 
\label{defcnk}
\end{eqnarray}
Let $c_{k}$ be $\sum_{n=1}^{N}c_{n,k}$.
The summation of $\zeta_{n}$  is written as
\begin{eqnarray}
\sum_{n=1}^{N}\zeta_{n} &=& 
\left \{
\begin{tabular}{ll}
$c_{\frac{N}{2}-1}\beta^{N-1} + 
\cal O \mit (\beta^{N}) $, & $N$ $=$ \mbox{even,}\\
$c_{N-1}\beta^{2N-1}+\cal O \mit (\beta^{2N}) $, & $N$ $=$ \mbox{odd.}
\end{tabular}
\right.
\end{eqnarray}
The details of the calculation of the  summation  are provided in the Appendix.
Using $c_{N/2-1}$ = $2N^2a_{N-1}$ and $c_{N-1}$ = $4N^2a_{2N-1}$,
the asymptotic behavior of the magnetic susceptibility at high temperatures is expressed as
\begin{eqnarray}
\chi
&\approx& 
-2\chi_{0} \frac
{\cos^2\left( \frac{\pi}{N}\right)}
{\sin\left( \frac{\pi}{N}\right)} N^2 a_{N-1}
\beta^{N-1}+\cal O \mit (\beta^{N}),
\hspace{5.5mm}\, N = \mbox{even,} 
\label{asybhighT1}
\\
\chi
&\approx& 
-4\chi_{0}\frac
{\cos^2\left( \frac{\pi}{N}\right)}
{\sin\left( \frac{\pi}{N}\right)} N^2 a_{2N-1}
\beta^{2N-1}+\cal O \mit (\beta^{2N}),
\hspace{3mm}\, N = \mbox{odd}.
\label{asybhighT2}
\end{eqnarray}
The results  show that the absolute value of the magnetic susceptibility decreases with the power laws 
$T^{-N+1}$ and $T^{-2N+1}$ for $N$ = even and odd, respectively.
Figure 4 shows the magnetic susceptibility for $N$ = 4, 5, and 6 in the log-log scale.
The circles shows the exact values, and   
the dashed lines are straight lines determined from Eqs. (\ref{asybhighT1}) and (\ref{asybhighT2}).
The susceptibility decreases more rapidly with size particularly for odd $N$. 

\begin{figure}[h]
\hspace{20mm}
\begin{center}
\includegraphics[bb= 0 0 260 210 , origin = c, clip,scale=1.0]{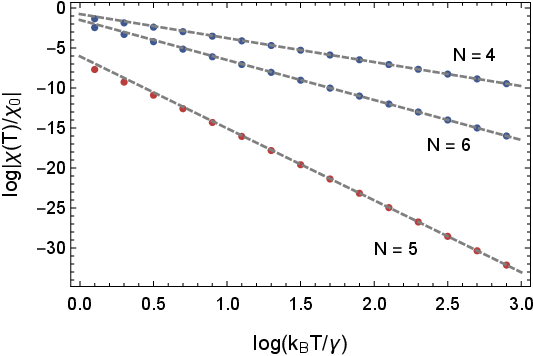}
\end{center}
\caption{Temperature dependence of the magnetic susceptibility normalized with $\chi_{0}$ at high temperatures in the double logarithm scale. Solid circles show the exact values, and the dashed lines indicate the asymptotes.}
\end{figure}

Because the free energy  is represented as a function of the magnetic flux density,
 the dependence of physical quantities other than magnetic susceptibility on the magnetic field $H$, can also be calculated. For example, 
 the magnetocaloric effect \cite{Oliveira2010} of a ring under adiabatic change can be calculated in terms of  the free energy as follows:
\begin{eqnarray}
\frac{\partial T}{\partial H} 
=\frac{T}{C_{H}}
\frac{\partial^2 F}{\partial T\partial B}.
\end{eqnarray}
where $C_{H}$ the heat capacity at constant magnetic field.


\section{Magnetic susceptibility of a belt}
Electron hopping  between two rings is possible in a belt.
We distinguish between the upper and lower rings with integers $k$ = 1 and 2, respectively.  Each site 
in the belt is labeled by $j$ ($1$ $\leq$ $j$ $\leq$ $N$) and $k$. The total number of sites is 2$N$. Unlike the rings in Section 2, $N$ must be even for the belt. The Hamiltonian of a belt is expressed as
\begin{eqnarray}
H_{\mbox{\tiny Belt}} &=& -\gamma\sum_{k=1}^{2}\sum_{j=1}^{N}
(e^{i\phi_{j}}c_{j+1,k}^{\dag}c_{j,k}+
e^{-i\phi_{j}}c_{j,k}^{\dag}c_{j+1,k}) \nonumber \\
&&
-\gamma^{\prime}\sum_{
\renewcommand{\arraystretch}{0.4}
\begin{tabular}{l}
\scriptsize{$j$ $=$ $1$} \\
\scriptsize{$j$ odd} \\
\end{tabular}
}^{N}
(c_{j,2}^{\dag}c_{j,1}+
c_{j,1}^{\dag}c_{j,2})
,
\label{Hamiltonian}
\end{eqnarray}
where $-\gamma^{\prime}$ is the transfer integral between the odd sites of the upper and lower rings. We assume that the transition between even sites is prohibited.

The  energy eigenvalues of $H_{\mbox{\tiny Belt}}$ normalized by $\gamma$ are given by
\begin{eqnarray}
\lambda_{n,l,N}(\alpha,t) &=&
\frac{1}{2}\left(
-p_{l}t+q_{l}
\sqrt{
t^2+
16\cos^2
\left[
\frac{2n\pi}{N}
+\frac{\alpha}{2}
\sin\left( \frac{2\pi}{N}\right) 
\right]
}
\right),
\nonumber \\
&& \hspace{60 mm} n = 1, 2, \ldots, N/2,
\end{eqnarray}
where $t$ = $\gamma^{\prime}/\gamma$, and 
$\{p_{l},q_{l}\}$ = $\{-1,-1\}$, $\{-1,1\}$, $\{1,-1\}$, and $\{1,1\}$ for $l$ = 1, 2, 3, and 4, respectively
where $t$ = $\gamma^{\prime}/\gamma$, and 
$p_{l}$ =$1$ and $-1$ for $l$ = 1, and 2, respectively. The components of eigenstates labeled with $(l,j,k)$ and $p_{l}$,
$\psi_{n,l,j,k}$
are expressed as
\begin{eqnarray}
\psi_{n,l,j,1} &=& C_{n,l}v_{n,l,j},\\
\psi_{n,l,j,2} &=& p_{l}C_{n,l}v_{n,l,j+N},
\end{eqnarray}
where $C_{n,l}$ is a normalized constant  determined from the condition that
$\sum_{l,j}(|\psi_{n,l,j,1}|^2+|\psi_{n,l,j,2}|^2) $ = 1, and $v_{n,l,j}$ is defined by
\begin{eqnarray}
v_{n,l,j} &=& 
\left \{
\begin{tabular}{ll}
$\lambda_{n,l,N}(\alpha,t)e^{i\theta_{n,j}}$,
& $j\,= \mbox{odd}$, \\
$\lambda_{n,N}(\alpha)e^{i\theta_{n,j}}$,
& $j\,= \mbox{even}.$
\end{tabular} 
\right.
\end{eqnarray}
In contrast to the eigenstate presented in Section 2, these expressions of the eigenstates do not form an orthogonal basis but  they satisfy
the Schr\"{o}dinger equation as follows.
If $j$ is even, the state after applying $H_{\mbox{\tiny Belt}}$ to the eigenstate $\psi_{n}$ is written as
\begin{eqnarray}
[H_{\mbox{\tiny Belt}}\psi_{n}]_{l,j,k}&=&-\gamma(e^{i\phi_{j-1}}\psi_{n,l,j-1,k}
+e^{-i \phi_{j}} \psi_{n,l,j+1,k}), \nonumber \\
&=&-\gamma(C_{n,l}e^{i\phi_{j-1}} \lambda_{n,l,N}e^{i\theta_{n,k}}
+C_{n,l}e^{-i \phi_{j}}\lambda_{n,l,N}e^{i\theta_{n,k}}), \nonumber \\
&=&-\gamma\lambda_{n,l,N}(\lambda_{n,N}e^{i\theta_{n,k}}), \nonumber \\
&=&-\gamma\lambda_{n,l,N}\psi_{n,l,j,k}.
\end{eqnarray}
If $j$ is odd and $k$ = 1, we have
\begin{eqnarray}
[H_{\mbox{\tiny Belt}}\psi_{n}]_{l,j,1}&=&-\gamma(e^{i\phi_{j-1}} \psi_{n,l,j-1,1}
+e^{-i \phi_{j}}\psi_{n,l,j+1,1}
+t\psi_{n,l,j,2}),
\nonumber \\
&=&-\gamma(\lambda_{n,N}(\lambda_{n,N}e^{i\theta_{n,k}})
+p_{l}t\lambda_{n,l,N}e^{i\theta_{n,k}}),
\nonumber \\
&=&-\gamma\lambda_{n,l,N}\psi_{n,l,j,k}.
\end{eqnarray}
Here, to obtain the last equation, we used the following equation:
\begin{eqnarray}
\lambda_{n,l,N}^2+p_{l}t\lambda_{n,l,N}-\lambda_{n,N}^2=0.
\end{eqnarray}
Similarly,  the Schr\"{o}dinger equation is satisfied for 
the case where $j$ is odd and $k$ = 2. 

Figure 5 shows the normalized energy eigenvalues of the belts for $N$ = 4 and 6 at $t$ = 0.2. For $N$ = 4, the energy eigenvalues of $(n,l)$ = $(1,2)$ and $(1,4)$ are zero at $\alpha$ = 0, and this degeneracy remains even when the transfer of electrons between rings occurs.
\begin{figure}[h]
\hspace{20mm}
\begin{center}
\includegraphics[bb= 0 0 260 310 , origin = c, clip,scale=1.0]{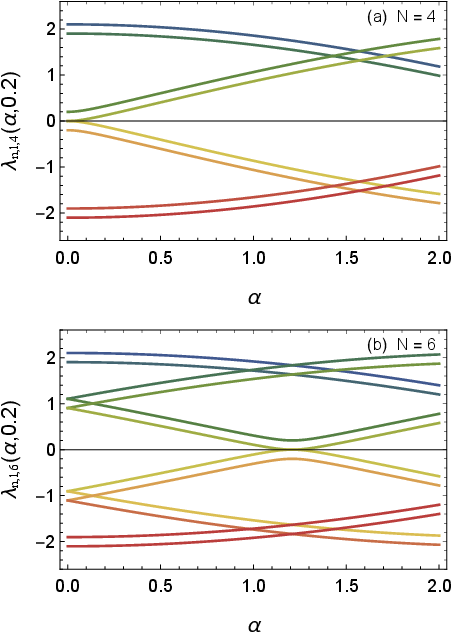}
\end{center}
\caption{ Energy eigenvalues normalized of belts with $\gamma$  for (a) 
$N$ = 4 and (b) $N$ = 6. The ratio of $\gamma^{\prime}$ to $\gamma$ is 0.2. }
\end{figure}
The first  and second derivatives of the normalized energy eigenvalues are, respectively, 
given by
\begin{eqnarray}
\left.\frac{\partial \lambda_{n,l,N}(\alpha,t)}{\partial \alpha}\right|_{\alpha =0} &=&
\frac{2p_{l}q_{l}\sin\left(\frac{2\pi}{N}\right)
\sin\left(\frac{4\pi n}{N}\right)
}
{\sqrt{t^2+8+8\cos\left(\frac{4\pi n}{N}\right)}}, \\
\left.\frac{\partial^2 \lambda_{n,l,N}(\alpha,t)}{\partial^2 \alpha}\right|_{\alpha =0} &=&
\frac{2p_{l}q_{l}\sin^2\left(\frac{2\pi}{N}\right)
\left \{
6+(8+t^2)
\cos\left(\frac{4\pi n}{N}\right)
+2\cos\left(\frac{8\pi n}{N}\right)
\right \}
}
{(t^2+8+8\cos\left(\frac{4\pi n}{N}\right))^{\frac{3}{2}}}.
\nonumber \\
\end{eqnarray}

Figure 6 shows the temperature dependence of the magnetic susceptibility of 
the belts with $t$ = 1 for $N$ = 4, 6, and 10. In contrast to Fig. 3, the susceptibility at absolute zero for $N$ = 4 is finite. We recall that the second term in the left-hand side of Eq. (\ref{chiesum1}) causes the divergence for rings. In the case of the belt with $N$ = 4 and $t$ = 0, the derivative of the energy eigenvalues with $n$ = 2 and 3 near $\alpha$ = 0 is not zero (see Fig. 2), whereas, if $t$ is not zero, the first derivative of the energy eigenvalues of the belt with $N$ = 4 for $(n,l)$ = $(1,2)$ and $(1,4)$ converges to zero (see Fig. 5), as  $\alpha$ approaches zero.  Thus, the second term vanishes in the limit of $\alpha$ $\rightarrow$ 0, and the divergence does not occur. This  finiteness of magnetic susceptibility for belts  holds for any size, if $N$ is a multiple of four.

\begin{figure}[h]
\hspace{20mm}
\begin{center}
\includegraphics[bb= 0 0 260 180 , origin = c, clip,scale=1.0]{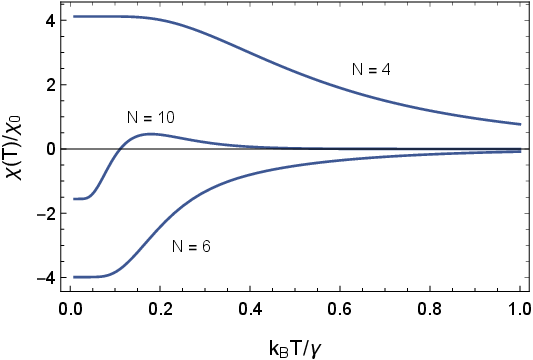}
\end{center}
\caption{Temperature dependence of the magnetic susceptibilities of the belts normalized with $\gamma$ for 
$N$ = 4, 6, and 10. }
\end{figure}

Interestingly, the sign of susceptibility for $N$ = 10 changes from negative to positive as the temperature increases.
Figure 7(a) shows the phase diagram of the sign in a plane of the transfer integral $t$ and the normalized temperature for $N$ = 10.
The sign is always negative for small $t$ values, independent of temperature. Above a critical value of $t$ near 0.76, the sign changes from negative to positive as the temperature increases from absolute zero. 
Conversely, the sign of the belt with $N$ = 4 changes from positive to negative as the temperature increases from absolute zero as shown in Fig. 7(b). 

\begin{figure}[h]
\hspace{20mm}
\begin{center}
\includegraphics[bb= 0 0 280 320 , origin = c, clip,scale=1.0]{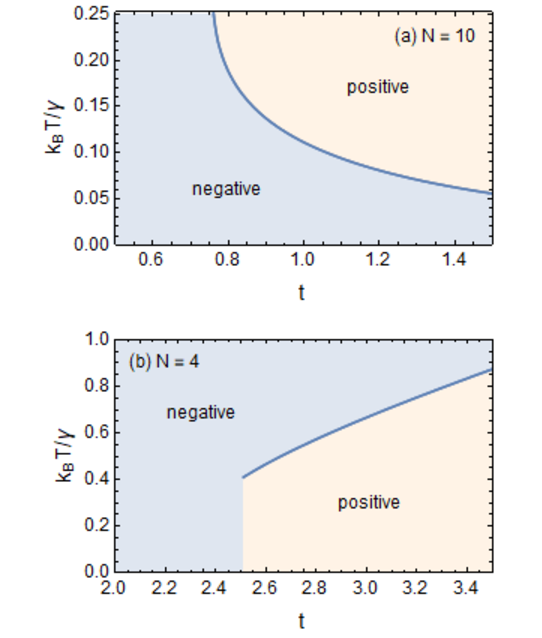}
\end{center}
\caption{Phase diagram of the belts divided by the sign of the magnetic susceptibility for (a) $N$ = 10 and (b) $N$ = 4. }
\end{figure}

\section{Conclusion}
We showed that the magnetic properties of  carbon nanobelts strongly depend on the divisibility of the number of sites by four.
If the number of sites in the ring is a multiple of four, the magnetic susceptibility diverges as the temperature approaches  absolute zero. This singularity occurs when the following conditions 
are satisfied. First, the energy eigenvalues converges to zero as the magnetic field decreases to zero. 
Second, their first derivative with respect to the magnetic field at $B$ = 0 is finite.
If the chemical potential $\mu$ is not zero, the first condition should be written as the condition that the energy eigenvalues  converge to the chemical potential as the magnetic field decreases.
As we assume that  $\mu$ = 0, the divergence occurs for $N$ = 4.
The number of distinct energy eigenvalues increases as the number of sites increases. 
Thus, the divergence can occur for various values of the chemical potential as $N$ increases. For example, if $\mu$ = $\gamma$,
the divergence occurs for $N$ = 6 (see Fig. 2(c)).

The energy eigenvalues of the belts change depending on the parameter $t$, which is a ratio of 
the transfer integral along the circumference and width directions.
For a belt consisting of two carbon rings with $N$ = 4, 
the derivative of energy eigenvalues at zero becomes zero by the presence of the transfer integral along the width direction. It follows that the  divergence at absolute zero does not occurs. However, this does not imply that no divergence  occur for $t$ $\neq$ 0.
The magnetic susceptibility of the belt with $N$ = 6  diverges if the parameter $t$ satisfying $\gamma \lambda_{n,N}(0,t)$ = $\mu$ exists.

This study focuses primarily on the interaction between electrons and an external magnetic field, and the many-body effect of electrons on the magnetic properties is considered indirectly via Pauli's exclusion principle. To describe the electron systems in the carbon nanobelt more accurately, the interactions between electrons should be considered. The most important interaction is likely the Coulomb interaction between electrons, which may be considered using the Hubbard model, whose interaction is described by 
$c_{j\uparrow}^{\dag}c_{j\uparrow}c_{j\downarrow}^{\dag}c_{j\downarrow}$, where $\uparrow$ and $\downarrow$ denote spin up and spin down, respectively. For graphene nanoribbons\cite{Kumar2020,Lou2024}, more realistic transport results have been obtained using the Hubbard model \cite{Hancock2010}. 

Various carbon nanobelts and rings shapes have been proposed and synthesized. 
For example, M\"{o}bius carbon nanobelts, with a twisted structure are smoothly connected at the front and back \cite{Segawa2022,Heine2020}. Their magnetic properties may be significantly different from that the belts considered in this study.  


\begin{flushleft}
{\bf Acknowledgments}
\end{flushleft}

This research was supported by the Ministry of Education, Culture, Sports, Science and Technology, through a Grant-in-Aid for Scientific Research(C), MEXT KAKENHI Grant Number 21K04895.

\appendix*
\section{Derivations of the Taylor series}
The coefficient of the Taylor series of $G_{1}$ is expressed as
\begin{eqnarray}
a_{k} = \left. \frac{d^k G_{1}(x)}{dx^{k}} \right|_{x=0}.
\end{eqnarray}
The $k$-th derivative of $G_{1}(x)$ is given by
\begin{eqnarray}
\frac{d^k G_{1}(x)}{dx^{k}}
&=& \sum_{j=1}^{k} (-1)^j\frac{T_{k,j}e^{-jx}}{(1+e^{-x})^{j+1}},
\end{eqnarray}
where
\begin{eqnarray}
T_{k, j} &=& 
\sum_{i=0}^{j}
(-1)^{j - i}
\binom{j}{i}
i^{k}.
\end{eqnarray}
Thus, $a_{k}$ is expressed as
\begin{eqnarray}
a_{k} &=&
-\frac{1}{k!}\sum_{j=1}^{k}\sum_{i=1}^{j}
\frac{(-1)^i}{2^{j + 1}} 
\binom{j}{i}
i^k.
\label{ak}
\end{eqnarray}
Comparing $G_{1}(x)$ with $G_{2}(x)$, 
we obtain the following relation:
\begin{eqnarray}
G_{2}(x) &=& 2\frac{dG_{1}(2x)}{dx}.
\end{eqnarray}
Accordingly, the coefficient $b_{2k}$ can be expressed using 
$a_{2k+1}$ as  Eq. (\ref{b2k})
The coefficient $c_{k}$ defined in Eq. (\ref{defcnk}) is written as
\begin{eqnarray}
c_{k,N} &=&2^{2k+2}a_{2k+1}d_{k+1,N}
+
b_{2k}d_{k+1,N}
-
b_{2k}
d_{k,N},
\label{ck}
\end{eqnarray}
where
\begin{eqnarray}
d_{k,N} &\equiv&\sum_{n=1}^{N}\cos^{2k}(\xi_{n}).
\end{eqnarray}
We define $h_{k,N}$ as
\begin{eqnarray}
h_{k,N} &\equiv&
\frac{N}{4^{k}}
\binom{2k}{k}.
\label{hk}
\end{eqnarray}
If $N$ is even and $k$ $<$ $N/2$, $d_{k,N}$ is given by $h_{k,N}$.
If $N$ is even and $k = N/2$, an additional term is required for $h_{k,N}$, 
and $d_{N/2,N}$ 
is expressed as  
\begin{eqnarray}
d_{\frac{N}{2},N} &=& 
h_{\frac{N}{2},N}
+\frac{N}{2^{N-1}}, \hspace{3mm} N = \mbox{even}.
\label{dhe}
\end{eqnarray}
These change determine the exponent of  decay for the magnetic 
susceptibility at high temperatures.
Similarly, if $N$ is odd and $k$ $<$ $N$, $d_{k,N}$ is given by $h_{k,N}$.
If $N$ is even and $k = N$,
\begin{eqnarray}
d_{N,N} &=& 
h_{N,N}
+\frac{N}{2^{2N-1}}, \hspace{3mm} N = \mbox{odd}.
\end{eqnarray}
Let $N$ be even.
Substituting Eqs. (\ref{ak}), (\ref{b2k}), and (\ref{dhe})  into the left-hand side of 
(\ref{ck}), we observe that $c_{k}$ is zero for $k$ $<$ $N/2-1$.
Thus, the first nonzero coefficient, $c_{N/2-1}$ is given by 
\begin{eqnarray}
c_{\frac{N}{2}-1,N} &=& 
2^{N}Na_{N-1}
\left\{
(1-N)h_{\frac{N}{2}-1,N}+Nh_{\frac{N}{2},N}+\frac{N}{2^{N-1}}
\right\}, \\
&=&2N^2a_{N-1}.
\end{eqnarray}
Similarly, if $N$ is odd, $c_{k}$ is zero for $k$ $<$ $N-1$, and 
$c_{N-1}$ $=$ $4N^2a_{2N-1}$.

\end{document}